\documentclass[11pt,a4paper]{article}

\usepackage{jcappub}
\usepackage[latin1]{inputenc}
\usepackage{graphicx}
\usepackage{epsfig}
\usepackage{subfigure}
\usepackage{color}
\usepackage{comment}
\usepackage{slashed}

\def\beq{\begin{equation}}
\def\eeq{\end{equation}}
\def\baq{\begin{eqnarray}}
\def\eaq{\end{eqnarray}}

\newcommand{\be}{\begin{equation}} 
\newcommand{\ee}{\end{equation}}
\newcommand{\bea}{\begin{eqnarray}} 
\newcommand{\eea}{\end{eqnarray}}

\newcommand{\bmp}{\noindent\begin{minipage}{16cm}}
\newcommand{\emp}{\end{minipage}\vskip 7mm} 
\def\lsim{\mathrel{\raise.3ex\hbox{$<$\kern-.75em\lower1ex\hbox{$\sim$}}}}
\def\gsim{\mathrel{\raise.3ex\hbox{$>$\kern-.75em\lower1ex\hbox{$\sim$}}}}

\newcommand{\intron}[1]{}

\newcommand{\Feyn}[1]{#1\kern-0.59em/}

\title{Observational signatures of Higgs inflation}

\author[a,b]{Vera-Maria Enckell,}
\author[a,b]{Kari Enqvist,}
\author[c,b]{and Sami Nurmi}

\affiliation[a]{Department of Physics, University of Helsinki \\
                      P.O.~Box 64, FI-00014, Helsinki, Finland}
\affiliation[b]{Helsinki Institute of Physics, \\
                      P.O.~Box 64, FI-00014, Helsinki, Finland}   
\affiliation[c]{Department of Physics, University of Jyv{\"a}skyl{\"a}, \\
                      P.O.Box 35 (YFL), FI-40014 University of Jyv{\"a}skyl{\"a}, Finland}
                            
\emailAdd{vera-maria.enckell@helsinki.fi}
\emailAdd{kari.enqvist@helsinki.fi}
\emailAdd{sami.t.nurmi@jyu.fi}

\abstract{We investigate the dependency of Higgs inflation on the non-renormalisable matching between the low energy Standard Model limit and the inflationary regime at high energies.  We show that for the top mass range $m_t \gtrsim 171.8$ GeV the scenario robustly predicts the spectral index $n_s \simeq 0.97$ and the tensor-to-scalar ratio $r\simeq 0.003$. The matching is however non-trivial, even the best-fit values $m_h=125.09$ GeV and $m_t=173.21$ GeV require a jump $\delta \lambda \sim 0.01$ in the Higgs coupling below the inflationary scale. For $m_t\lesssim  171.8$ GeV, the matching may generate a feature in the inflationary potential. In this case the predicted values of $n_s$ and $r$ vary but the model is still falsifiable. For example, a detection of negative running of spectral index at level $\alpha_s \lesssim -0.01$ would rule out Higgs inflation.

}

\begin{document}
\maketitle

\section{Introduction}

The Standard Model (SM) Higgs could be the inflaton field explaining the origin of structure in the universe. 
The Higgs inflation scenario \cite{Bezrukov:2007ep} assumes a large non-minimal curvature coupling $\xi R h^2$ for the SM Higgs. This ensures the required flat potential during inflation assuming that higher order curvature terms are suppressed. The suppression is a non-trivial constraint on ultraviolet physics \cite{Burgess:2014lza}. However  the setup is self-consistent as all the energy scales are below the perturbative unitarity cutoff during inflation \cite{Bezrukov:2010jz,Bezrukov:2014bra}.  

However, even in the absence of higher order curvature operators, the inflationary potential is not uniquely determined by the measured SM parameters and the non-minimal coupling $\xi$ \cite{Bezrukov:2014bra,Bezrukov:2014ipa,Fumagalli:2016lls}. This is because the non-renormalisable setup which does not allow straightforward running of the SM couplings up to the inflationary scale. The running of SM couplings can be computed perturbatively around the electroweak vacuum $h=246$ GeV up to energies $\mu\sim M_P/\xi$. In the inflationary regime $h\gtrsim M_P/\xi$ the Higgs potential is approximatively shift symmetric which enables perturbative treatment of radiative corrections also over the scales $M_P/\xi \lesssim \mu\lesssim M_P/\sqrt{\xi}$ \cite{Bezrukov:2009db, Bezrukov:2010jz,George:2013iia}. The matching of these two regimes is however obscured by manifestly non-renormalisable physics at intermediate scales \cite{Bezrukov:2014bra,Bezrukov:2014ipa,Fumagalli:2016lls}.  
 
Assuming the Higgs inflation indeed took place in the early universe, the system of SM coupled to gravity should have an ultraviolet completion which eventually determines the matching \cite{Bezrukov:2009db, Bezrukov:2014bra,Bezrukov:2014ipa, Barbon:2015fla}. Without the exact solution at hand, however, we are led to parameterise the effects of the unknown non-renormalisable physics. Here we follow \cite{Bezrukov:2014bra,Bezrukov:2014ipa} and model its impact by introducing two effective parameters, $\delta\lambda$ and $\delta y_t$, which represent unknown jumps in the Higgs and top Yukawa couplings at the intermediate scale $\mu\sim M_P/\xi$.  These parameters together with the perturbative running at low and high energy regimes then connect the measured SM couplings to the effective Higgs potential during inflation.  

There are several consistency constraints which restrict the viable choices of parameters $\delta\lambda$ and $\delta y_t$.  First, the jump $\delta\lambda$ cannot in general be chosen zero. This is because the Higgs coupling runs negative before the inflationary scale unless the top quark is significantly below its best fit value $m_t\simeq 173.21$ GeV \cite{Degrassi}.  Inflation then requires the coupling to jump positive before inflation which necessitates a large enough positive $\delta\lambda$ \cite{Bezrukov:2014ipa}. 
Second, the required relaxation of the inflationary Higgs condensate into the SM vacuum during reheating further constraints $\delta\lambda$ and $\delta y_t$. The reheating constraints are non-trivial if the Higgs coupling runs negative and generates a negative energy minimum between the inflationary plateau and the SM vacuum.
Third, the primordial perturbations generated during inflation should have the observed properties which also constrains the matching parameters. Specific choices of $\delta\lambda$ and $\delta y_t$ may generate a feature in the inflationary potential which in the extreme case becomes an inflection point \cite{Hamada:2014iga, Bezrukov:2014bra,Bezrukov:2014ipa,Fairbairn:2014nxa}. 

In this work we investigate in detail the impact of matching parameters in Higgs inflation. By varying the Higgs and top masses within their experimental two-sigma limits and scanning over a wide range of $\delta\lambda,\delta y_t$ and $\xi$, we establish a systematical picture of the full range of possible observational signatures  in the scenario. We show that for top mass values $m_t \gtrsim 171.8 $ GeV, the observed amplitude of primordial perturbations ${\cal P}_{\zeta}=2.138 \times 10^{-9}$ \cite{Ade:2015xua} can be obtained only for parameter combinations which yield the spectral index and tensor to scalar ratio consistent with the standard predictions of Higgs inflation, $n=0.97$ and $r_T=0.003$ \cite{Bezrukov:2007ep,Barvinsky:2008ia}. For $m_t\lesssim 171.8$ GeV the observed amplitude can also be obtained for matching parameters which generate a feature in the inflationary potential and still allow for reheating into the SM vacuum. In this case $n_s$ and $r_T$ can vary over a broad range of values \cite{Hamada:2014iga, Bezrukov:2014bra,Bezrukov:2014ipa}. However, we show that the scenario is still falsifiable as the predictions for the spectral index $n_s$, its running $\alpha_s$ and the tensor-to-scalar ratio $r_T$ always satisfy a consistency relation. Combining the results, we discuss in detail how the requirement of successful Higgs inflation followed by reheating into the SM vacuum constrain the viable values of $\delta\lambda$ and $\delta y_t$ for different choices of the Higgs and top masses and the non-minimal coupling. 

The paper is organised as follows. In Section \ref{sec:setup}  we summarise the setup for Higgs inflation and introduce the matching parameters. In Section \ref{sec:potential} we discuss the form of the effective Higgs potential during inflation and its dependence on the matching. In Section \ref{sec:conditions} we discuss the consistency conditions on matching parameters implied by reheating and the end of inflation. In Section \ref{sec:predictions} we analyse the observational signatures of the model and investigate how the predictions for a given choice of SM parameters depend on the matching. Finally, in Section \ref{sec:conclusions} we present our conclusions.

\section{The setup for Higgs inflation}
\label{sec:setup}

Including the non-minimal Higgs coupling to spacetime curvature the Lagrangian for gravity and Standard Model fields reads 
\beq
\label{Lnonminimal}
{\cal L} =\frac{ M_P^2}{2}R + \frac{\xi}{2} h^2 R + {\cal L}_{\rm SM}\ . 
\eeq
Here the gauge-invariant operator $h^2 = 2 \Phi^{\dag}\Phi$ denotes the length of the Higgs doublet $\Phi$ and in the following we will call $h$ the Higgs field. The non-minimal coupling is inevitably generated by radiative effects and it has a major role in explaining the vacuum stability even if the Higgs would not act as the inflaton \cite{Espinosa:2007qp,Herranen:2014cua,Lebedev}. In addition to the non-minimal coupling a generic effective Lagrangian is expected to contain all possible higher order terms. In Higgs inflation the couplings of higher order operators are required to be small and $\xi$ should be large \cite{Bezrukov:2007ep} which are non-generic conditions on the ultraviolet physics \cite{Burgess:2014lza}. However, the setup is self-consistent as the perturbative unitarity is preserved during inflation without any need to include higher order terms  \cite{Bezrukov:2010jz,Bezrukov:2014bra}.  

It is convenient to make a conformal transfomation in Eq. (\ref{Lnonminimal}) such that ${g}_{\mu\nu} \rightarrow \Omega^2 g_{\mu\nu}$ with $\Omega^2 = 1+ \xi h^2/M_P$. In this so called Einstein frame the gravitational part of the Lagrangian takes the form of Einstein gravity. The scalar kinetic term can be brought to canonical form in this frame by a suitable redefinition of the Higgs field $\chi(h)$ and a rescaling of the fermions by $\psi \rightarrow \psi/\Omega^{3/2}$.  In terms of the canonically normalised Einstein frame quantities the parts of the Lagrangian (\ref{Lnonminimal}) including gravity and couplings of the Higgs boson and top quark are given by \cite{Bezrukov:2014bra} 
\beq
\label{Leinstein}
{\cal L} =\frac{1}{2}M_P^2R + \frac{1}{2}D^{\mu} \chi D_{\mu}\chi - \frac{\lambda}{4} F^4(\chi) + i\bar{\psi}_t\Feyn{D}\psi_t+ \frac{y_t}{\sqrt{2}}F(\chi)\bar{\psi}_t\psi_t\ . 
\eeq
Here $F(\chi)$ reduces to $\chi\simeq h$ for small field values and approaches constant for large field values  
\beq
\label{F}
F(\chi) = \frac{h}{(1+\xi h^2/M_P^2)^{1/2}} =
\begin{cases}
\chi & \qquad, ~\chi \ll M_P/\xi\\
\frac{M_P}{\sqrt{\xi}}\left(1-{\rm e}^{-\sqrt{2/3}\; \chi/M_P}\right)^{1/2} & \qquad, ~\chi \gg M_P/\xi
\end{cases}
\ .
\eeq

In the small field limit $\chi\ll M_P/\xi$ the action (\ref{Leinstein}) reduces to SM in flat space. Radiative corrections can be computed perturbatively up to the unitarity cutoff $M_P/\xi$ where non-renormalisable gravitational interactions can no longer be neglected. On the other hand, in the inflationary regime, $\chi\gg M_P/\xi$, the effective potential has an approximative shift symmetry and the unitarity cutoff is pushed up to $M_P/\sqrt{\xi}$ \cite{Bezrukov:2014bra,Bezrukov:2014ipa}. This warrants perturbative treatment of radiative corrections also over scales $M_P/\xi \lesssim \mu \lesssim M_P/\sqrt{\xi}$. 

However, in the intermediate regime $\mu \sim M_P/\xi$, non-renormalisable operators are not suppressed and the perturbative analysis loses its validity. This leads to ambiguity in matching the low energy SM limit and the inflationary high energy regime  \cite{Bezrukov:2014bra,Bezrukov:2014ipa}. Therefore, even if there would be no higher order curvature corrections to Eq. (\ref{Lnonminimal}) the inflationary potential is not uniquely determined by the measured SM parameters and the non-minimal coupling $\xi$ alone.

\subsection{Matching the low and high energy regimes}

Here we follow the approach of \cite{Bezrukov:2014bra,Bezrukov:2014ipa} and postulate that the system of SM coupled to gravity has an ultraviolet completion described by Eq. (\ref{Leinstein}). The matching of the SM limit $\chi\ll M_P/\xi$ and the inflationary regime $\chi\gg M_P/\xi$ should then be determined by the exact non-perturbative solution. As the form of the postulated solution is not known, however, the matching is undetermined. Here we adopt a phenomenological approach and parameterise 
our ignorance of the non-perturbative solution.  We then proceed to systematically investigate how predictions of the Higgs inflation depend on the unknown physics and how the matching implied by successful Higgs inflation differs for different choices of SM parameters. 
 
Following  \cite{Bezrukov:2014bra,Bezrukov:2014ipa} we introduce effective matching parameters to connect the low and high energy solutions for the running SM of couplings. The beta functions can be computed perturbatively in both limits. However, their solutions in the high energy regime contain a priori arbitrary constants of integration. Matching the two regimes amounts to determining these constants.  For example, cancelling one-loop divergences in the effective Higgs coupling $\lambda$ requires adding a counterterm to Eq. (\ref{Leinstein}) of the form \cite{Bezrukov:2014ipa}
\beq
\delta {\cal L} =  \left(-\frac{2}{\epsilon}\frac{9\lambda^2}{64\pi^2} -\gamma + {\rm ln}(4\pi)+\delta \lambda\right)\left(F'{}^2+\frac{1}{3}F''F^2\right)^2F^4 \ ,
\eeq
where $\epsilon = 4-d$. This vanishes for $\chi\gg M_P/\xi$ and for $\chi\ll M_P/\xi$ it reduces to the operator $F^4$.  Consequently, the effective coupling $\lambda(\mu) F^{4}$ jumps by an arbitrary finite amount $\delta\lambda$ when moving from small field values to the inflationary regime. The top Yukawa $y_t$ and weak gauge couplings $g,g'$ may feature similar jumps which stem from mass corrections proportional to $F'$ and $F''$. In our analysis we will include only the jumps $\delta\lambda$ and $\delta y_t$ for the Higgs and top couplings. 

We choose to match the low and high energy regimes at the point, $\chi_1$, where $F(\chi)$ given by Eq. (\ref{F}) switches from one branch to another  
\beq
\label{chi1}
\chi_1=\frac{M_P}{\sqrt{\xi}}\left(1-{\rm e}^{-\sqrt{2/3}\; \chi_1/M_P}\right)^{1/2}~.
\eeq
In the following we will always choose the renormalisation scale $\mu$ equal to the top quark mass. This amounts to approximatively minimising the leading logarithmic corrections $\delta V \sim m_t^4{\rm ln}(\mu^2/m_t^2)$ to the effective potential. With this choice, the matching scale $\mu_1=\mu(\chi_1)$ is given by 
\beq
\label{mu1}
\mu_1=\frac{y_t(\mu_1)\chi_1}{\sqrt{2}}~.
\eeq

The running of couplings in the low energy limit $\mu\ll \mu_1$ is determined by the usual SM beta functions in flat space, see e.g. \cite{Degrassi}. In the high energy regime $\mu\gg \mu_1$ the Higgs potential settles to a constant and the system is formally analogous to chiral electroweak theory \cite{Bezrukov:2009db,Dutta:2007st}. In matching the two regimes we allow for jumps in the Higgs and top Yukawa couplings 
\baq
\label{jumps}
\lambda(\mu_1)&=&\lambda^{\rm SM}(\mu_1) + \delta\lambda ,\\\nonumber
y_{t}(\mu_1)&=&y_t^{\rm SM}(\mu_1) + \delta y_t\ .
\eaq
Here $\lambda^{\rm SM}(\mu)$ and $y_t^{\rm SM}(\mu)$ denote solutions of the SM renormalisation group equations and we evaluate all couplings in the $\overline{MS}$ scheme and Landau gauge. We solve them to next to next to leading order precision using the code available at \cite{bezrukoodi}.  We treat the jumps $\delta \lambda$ and $\delta y_t$ as a priori free parameters constrained only by the requirement of perturbativity, $|\lambda(\mu_1)|<1$ and $|y_t(\mu_1)|<1$. Note that the effective potential and couplings are not directly physical but the inflationary perturbations are. For a different choice of gauge and the renormalisation scheme, the effective potential and also the values of $\delta \lambda$ and $\delta y_t$ leading to the same observables would in general be different. Here we stick to one choice and scan over all possible values of the matching parameters. Hence, our results for the inflationary observables are in this regard fully general.

The running in the high-energy regime $\mu>\mu_1$ is solved from the one-loop chiral beta functions which read 
\baq
\label{betachiral}
16 \pi^2 \beta_{\lambda} &=& -6y_t^4+ \frac{3}{8}(2g^4+(g'{}^2+g^2)^2)+\lambda(-3g'{}^2-6g^2+12y_t^2)\ ,\\\nonumber
16 \pi^2\beta_{y_t} &=& y_t\left(-\frac{17}{12}g'{}^2- \frac{3}{2}g^2-8g_3^2+3y_t^2\right)\ ,\\\nonumber
16 \pi^2\beta_{g_3} &=&-7g_3^2\ , \\\nonumber
16 \pi^2\beta_{g'} &=&\frac{27}{4} g'{}^2\ , \\\nonumber
16 \pi^2\beta_{g} &=&-\frac{13}{4} g^2 \ .
\eaq
The initial conditions for $\lambda$ and $y_t$ are set by  Eq. (\ref{jumps}) at the matching scale $\mu=\mu_1$. We assume that there are no corresponding features in the gauge coupling RGEs. Their initial conditions are given by $g_{i}(\mu_1)=g_{i}^{\rm SM}(\mu_1)$ and we use the best fit values as the input for their SM running. The couplings in the inflationary regime thus depend on the set of five parameters $(m_h,m_t,\xi,\delta\lambda, \delta y_t)$. 

In Eq. (\ref{betachiral}) the last term of $\beta_{\lambda}$ is suppressed compared to the first two due to the smallness of $\lambda$ and we will neglect it. To the same precision we can also neglect the running of the non-minimal coupling $\xi$ \cite{Bezrukov:2009db} and we will hence treat $\xi$ as a constant. The change in its value over the inflationary energies is anyway negligible and its value at low energies is of little interest as there are no useful experimental bounds for its value there. 

\section{The inflationary potential}
\label{sec:potential}

For $\chi \gg v = 246 $ GeV the renormalisation group improved tree-level effective potential can be approximated by 
\beq
\label{potentiaali}
V (\chi)= \frac{\lambda(\mu)}{4}F^4(\chi)~,\qquad \mu=\frac{y_t (\mu)F(\chi)}{\sqrt{2}}.
\eeq
Here the running of the coupling $\lambda(\mu)$ from the low energy SM limit up to inflationary energies is determined by Eqs. (\ref{jumps}) and (\ref{betachiral}). We have chosen the renormalisation scale $\mu$ such that leading logarithmic corrections to the potential $\delta V \sim m_t^4{\rm ln}(\mu^2/m_t^2)$ are small and can be neglected. While the full effective potential does not depend on the choice of $\mu$, the error in its perturbative approximation in general does. 

The exponential form of Eq. (\ref{F}) drives the potential nearly constant  $V\simeq \lambda M_P^4/{4\xi^2}$  for large field values. This  makes inflation possible in the regime $\chi\gtrsim M_P$  \cite{Bezrukov:2007ep}. The amplitude of primordial perturbations, the spectral index, its running and the tensor-to-scalar ratio are given by the standard single field slow roll expressions. 
\baq
\label{SRobs}
{\cal P}_{\zeta}&=&\frac{M_P^{-2}}{2\epsilon}\left(\frac{H}{2\pi}\right)^2~,\qquad n_{s}-1\equiv \frac{{\rm d} {\cal P}_{\zeta}}{{\rm d}{\rm ln}k} =2\eta-6\epsilon~,\\\nonumber
\alpha_{s}&\equiv& \frac{{\rm d} n_s}{{\rm d}{\rm ln}k}= -2\Xi -24\epsilon^2+16\epsilon\eta~,\qquad r_{T}=16\epsilon~.
\eaq
The slow roll parameters are defined by 
\beq
\label{SRparam}
\epsilon=\frac{M_P^2}{2}\left(\frac{V'}{V}\right)^2~,\qquad \eta=M_P^2\frac{V''}{V}^2~,\qquad \Xi = M_P^4 \frac{V'V'''}{V^2}~.
\eeq
The third slow roll parameter is commonly denoted by $\xi$ but here we use $\Xi$ to distinguish it from the non-minimal coupling. All the quantities are evaluated at the horizon crossing of observable modes. Throughout this work we assume this corresponds to $N_{\rm CMB}=60$ e-folds before the end of inflation. 

The first and second derivatives of the Higgs potential Eq. (\ref{potentiaali}) are given by 
\baq
\label{V'}
V' &=& \frac{F^3 F'}{4}\left(4 \lambda +\beta_{\lambda}\right)\\
\label{V''}
V'' &=& \frac{(F^3 F')'}{F^3 F'} V'+ \frac{F^2 F'{}^2}{4}\left(4\beta_{\lambda}+\frac{d \beta_{\lambda}}{d {\rm ln}\mu}\right).
\eaq
Generically one expects that $\beta_{\lambda}$ and its derivative in Eqs. (\ref{V'}) and (\ref{V''}) can be neglected during inflation as they are suppressed by higher powers of couplings. This corresponds to the standard Higgs inflation \cite{Bezrukov:2007ep} where the amplitude ${\cal P}$ depends on the ratio $\lambda/\xi^2$ and the number of e-folds  $N_{\rm CMB}$ left at the horizon crossing of observable modes and all other quantities in Eq. (\ref{SRobs}) depend on $N_{\rm CMB}$ only. 

The situation is different in the non-generic parameter regime where $\lambda ={\cal O}(\beta_{\lambda})$ at the inflationary scale. In this case the running of $\lambda$ may lead to accidental suppression of the first and second derivatives \cite{Hamada:2014iga,Bezrukov:2014bra,Bezrukov:2014ipa}. Consequently, the inflationary potential may have a feature $V''\simeq 0$ or even an inflection point  $V'\simeq V''\simeq 0$. The spectrum of perturbations will then sensitively depend on the matching parameters. It should be noted that cancellation between parametrically  different quantities Eqs. (\ref{V'}) and (\ref{V''}) is not generic. For example, inflection point configurations require $\lambda = {\cal O}(\beta_{\lambda}^2/y_t^2)= {\cal O}(10^{-6})$ which implies fine-tuning of $\delta\lambda$ and $\delta y_t$ to the level of sixth digits  \cite{Bezrukov:2014bra,Bezrukov:2014ipa}. Here we leave questions of naturalness aside, and simply aim to systematically investigate all possible observational signatures of the Higgs inflation and to clarify their dependence on the matching. Note that with the two free parameters $\delta\lambda$ and $\delta y_t$ it is not possible to simultaneously have accidental suppression of more than two independent derivatives of the potential. We therefore neglect the second derivatives $d^2\beta_{\lambda}/d {\rm ln}\mu^2$ which would appear in $V'''$ and only include terms up to  $d\beta_{\lambda}/d {\rm ln}\mu$.

The possible features in the inflationary potential can be conveniently classified in terms $\delta_{1,2}$ defined by  
\baq 
\label{delta1}
4 \lambda(\mu) +\beta_{\lambda}(\mu) &\equiv&\lambda(\mu) \delta_1(\mu)\\
\label{delta2}
4\beta_{\lambda}(\mu)+\frac{d \beta_{\lambda}(\mu)}{d {\rm ln}\mu}&\equiv& \beta(\mu) \delta_2(\mu)~.
\eaq 
For each set of SM parameters and the non-minimal coupling $\xi$, the choice of matching parameters $\delta\lambda$ and $\delta y_t$ uniquely determines $\delta_{1,2}(\mu)$ through Eqs. (\ref{jumps}) and  (\ref{betachiral}). The cases where $\delta_{1,2}={\cal O}(1)$ correspond to an almost featureless potential and the spectrum of produced perturbations is to a good approximation described by the standard results of Higgs inflation. The inflection point is generated if $\delta_{1}=\delta_{2}=0$ at some scale $\mu$ in the inflationary regime. The cases between these limits correspond to milder features in the potential which may still affect the observables.

 By scanning over $\delta\lambda$ and $\delta y_t$ and numerically computing the inflationary dynamics we will establish a systematical picture of the observational signatures which can be obtained for each choice of  SM parameters and $\xi$. In addition to their impacts on inflationary observables, the matching parameters are generically constrained also by the reheating stage after the end of inflation. This holds true even for the standard case of Higgs inflation with no features in the potential. We now turn to discuss
these consistency conditions and then combine them with the analysis of observable signatures in Section \ref{sec:predictions}.

\section{Consistency conditions for the matching}
\label{sec:conditions}

For best fit values of the SM parameters the Higgs coupling $\lambda(\mu)$ becomes negative at scales $\mu \gtrsim 10^{10}$ GeV \cite{Degrassi}.  In this case the Higgs inflation can be realised only if the non-renormalisable physics at the intermediate scales $\mu\sim M_P/\xi$ affects the running such that $\lambda$ becomes again positive in the inflationary regime $\mu\gtrsim M_P/\xi$. In the effective parametrisation (\ref{jumps}) this requires a jump $\delta\lambda$ from negative to positive values at $\mu=\mu_{1}$.  The inflationary regime $\chi\gtrsim \chi_1$ will then be separated from the electroweak vacuum by a negative energy minimum of the potential located at $\chi\sim \chi_1$ \cite{Bezrukov:2014ipa}. 

In the presence of the negative energy minimum, the relaxation of the inflationary Higgs condensate into the electroweak vacuum during reheating is a non-trivial condition. If the minimum is sufficiently deep $|V(\chi_1)|\lesssim \lambda_{\rm inf} {M_P}^4 / {\xi}^2 $, the Higgs may instead end up into the negative energy minimum. In this case reheating of the observed Universe would not be possible. In \cite{Bezrukov:2014ipa} it was proposed that thermal corrections could lift the negative energy minimum leading to successful reheating. This requires large enough reheating temperature which constrains the viable range of matching parameters.

\subsection{Thermal corrections and reheating}

After the end of inflation the Higgs becomes massive $V''\sim H^2$ and starts to oscillate around the minimum of its potential. The dominant decay channel of the coherently oscillating condensate is the non-perturbative production of weak gauge bosons \cite{Bezrukov:2008ut,Enqvist:2015sua}. They will further decay into quarks and leptons generating a thermal bath of SM particles. Detailed analysis of the process is beyond the scope of the current work. Here we follow \cite{Bezrukov:2014ipa} and investigate when thermal corrections can lift the negative energy minimum in the most optimal case where all the inflationary energy is instantaneously converted into the thermal bath. This represents the minimal, necessary condition for reheating into the electroweak vacuum.  As the actual decay is not instant and the energy transfer is not complete the actual sufficient condition could be even tighter. 

Thermal corrections to the effective Higgs potential are given by 
\be\label{thermal}
\Delta V(T,\chi)= T \sum_{i} \int\frac{d^3k}{(2\pi)^3 a^3}{\rm ln}(1\pm e^{-\beta (k^2/a^2 + m_i^2(T))})~,
\ee
where the plus sign corresponds to fermions and minus sign to bosons. Here we will include only the dominant contributions which arise from top quarks and weak gauge bosons. Their thermal masses are given by 
\begin{align}
  m_Z^2(T) &= \frac{g_1^2(\mu_g)+g_2^2(\mu_g)}{4}F^2(\chi),\\
  m_W^2(T) &= \frac{g_2^2(\mu_g)}{4}F^2(\chi), \\
  m_t(T) &= \frac{y_t(\mu_t)}{\sqrt{2}}F(\chi),
\end{align}
and we evaluate the couplings respectively at scales $\mu_g = 7T$ and $\mu_t = 1.8 T$ \cite{Kajantie:1995dw}. In computing the couplings we use the SM renormalisation group equations matched to chiral SM at $\mu_1$ according to Eqs. (\ref{jumps}) and (\ref{betachiral}).

If the Higgs coupling runs negative below the inflationary scale, the zero temperature potential has a negative energy minimum, $V_{\rm min}= V(\chi_{\rm min})$, between the electroweak vacuum and the inflationary regime. We require that the negative energy minimum is lifted by thermal corrections such that the global minimum of the potential corresponds to $\chi=0 $. This implies the condition 
\beq
\label{successfulreh}
\Delta V(T_{\rm reh}, \chi_{\rm min}) >  -V(\chi_{\rm min})~.
\eeq
We evaluate the thermal corrections using the highest possible reheating temperature $T_{\rm reh}$ which corresponds to instant reheating and full conversion of the inflationary energy into radiation 
\begin{equation} \label{Tformula}
T_{\rm reh} = \left(\frac{30 V_{\rm inf}}{g_*\pi^2}  \right)^{1/4}~.
\end{equation}
Here $g_* = 106.75$ is the effective number of degrees of freedom. For most choices of the matching parameters, the negative minimum is located at the matching scale $V_{\rm min}\simeq \lambda(\mu_1)\chi_1^4/4$ where the Higgs coupling jumps from negative to positive. It is however also possible that $\lambda$ does not jump all the way up to positive values at $\mu_1$. It may instead jump close to zero and cross zero at a higher scale (still below inflation). In this case the minimum is located somewhere between $\chi_1$ and the inflationary regime. To account for all possible forms of the potential, we compute the effective potential Eq. (\ref{potentiaali}) and determine the location $\chi_1$ of the negative minimum separately for each parameter set $(m_h,m_t,\xi,\delta \lambda,\delta y_t)$.

In the following we will take Eqs. (\ref{successfulreh}) and (\ref{Tformula}) as our criteria for successful reheating. The conditions translate into bounds on the effective jump parameters $\delta\lambda$ and $\delta y_t$ which will depend both on the low energy SM parameters $m_h$ and $m_t$ and on the non-minimal coupling $\xi$. Increasing the top mass or decreasing the Higgs mass leads to more negative values of the Higgs coupling and hence tighter constraints. The non-minimal coupling enters the problem through two opposite effects. Decreasing $\xi$ moves the matching scale $\mu_1\propto \xi^{-1}$ to higher values leading to more negative $\lambda(\mu_1)$. On the other hand, it increases the inflationary energy scale as $V_{\rm inf}\propto \xi^{-2}$. Which of the effects dominates depends on the parameters and will be investigated numerically below. 

We reiterate that  Eqs. (\ref{successfulreh}) and (\ref{Tformula}) only give the minimal necessary conditions for reheating. The actual sufficient conditions for the relaxation of the Higgs condensate into the electroweak vacuum could be tighter. Determining the sufficient conditions would require a careful investigation of the reheating stage and resonant decay of the Higgs condensate accounting for the dynamical backreaction of the generated bath of light SM particles. This is beyond the scope of our current work. Moreover, as the phenomenological matching,  Eqs. (\ref{jumps}) and (\ref{betachiral}), may not represent the actual form of the negative energy minimum there is little point in going beyond the order of magnitude estimates within this simplistic prescription.  

\subsection{End of inflation}

Depending on the choice of $\delta\lambda$ and $\delta y_t$, the potential may contain a local maximum in the regime $\mu > \mu_1$. A maximum between the field value $\chi_{\rm CMB}$ at horizon exit of observable modes and end of slow roll regime $\chi_{\rm end}$ would be fatal as there would be no graceful exit from the inflationary regime. A maximum before the observable range of e-folds, $\chi_{\rm max}>\chi_{\rm CMB}$ is in principle be allowed, see e.g. \cite{Fumagalli:2016lls}. It would however imply tuning of  initial conditions. The initial field value should be chosen in the range $\chi<\chi_{\rm max}$ or otherwise there would again be no graceful exit. 

Here we require that the initial conditions for inflation can be chosen arbitrarily in the regime $\chi > \chi_{\rm CMB}$ and that there is always a graceful exit from inflation. We therefore impose the condition 
\beq
\label{negslope}
V'(\chi) \leqslant 0~,\qquad {\rm for}~~\chi>\chi_{\rm end}. 
\eeq
Together with the successful reheating condition Eqs. (\ref{successfulreh}) and (\ref{Tformula}) this further constrains the choices of matching parameters. This restricts in particular the jumps in the top Yukawa coupling $\delta y_t$ which directly affects the running  of $\lambda(\mu)$ in the inflationary regime.   

The condition Eq. (\ref{negslope}) still allows for choices of $\delta\lambda$ and $\delta y_t$ which may generate a local maximum after the end of slow roll inflation. In that case we require $V(\chi_{\rm max})<V(\chi_{\rm end})$ such that relaxation to the electroweak vacuum is possible.

\section{Observational signatures }
\label{sec:predictions}

We now turn to analyse the observational predictions of Higgs inflation. Our goal here is to establish a systematical picture of the observational signatures and their dependence on the matching of the low and high energy regimes. The results will also demonstrate how each choice of $m_h,m_t$ and $\xi$ in general implies a different connection between the two regimes separated by perturbatively non-renormalisable physics. 

In terms of the parameterisation introduced in Eqs. (\ref{delta1}) and (\ref{delta2}), the standard Higgs inflation \cite{Bezrukov:2007ep} corresponds to parameter sets $(m_h,m_t,\xi, \delta\lambda, \delta y_t)$ which yield  $\delta_{1,2}={\cal O}(1)$. In this case the running of $\lambda$ has a negligible during inflation and the inflationary dynamics essentially corresponds to $R^2$ inflation \cite{Starobinsky:1980te}. The spectrum, spectral index, its running and the tensor-to-scalar ratio are then to good approximation given by    
\beq
\label{standardobs}
{\cal P}_{\zeta} =\frac{\lambda N_{\rm CMB}^2}{72 \pi^2\xi^2}~,\qquad  n_s =1-\frac{2}{N_{\rm CMB}}~,\qquad  \alpha_{s} = -\frac{2}{N_{\rm CMB}^2}~,\qquad  {r}_{T} =~ \frac{12}{N_{\rm CMB}^2}\qquad~, 
\eeq
which do not directly depend on the matching parameters. Setting $N_{\rm CMB}=60$ one obtains the well-known standard predictions of the Higgs inflation. 

On the other hand, the inflationary dynamics is strongly dependent on the matching for the choices of $\delta\lambda$ and $\delta y_t$ which yield $\delta_{1}\ll 1$ or $|\delta_{2}|\ll 1$.  An extreme example is the inflection point scenario \cite{Bezrukov:2014bra,Bezrukov:2014ipa} which corresponds to $\delta_{1}=\delta_{2}=0$ such that the first and second derivatives of the potential vanish locally. More generally, the inflationary potential may contain a milder feature where the second derivative vanishes locally $V''=0$ but the first derivative may be non-vanishing. The observational signatures then depend sensitively on $\delta\lambda$ and $\delta y_t$ which determine both the location of the feature and the form of the potential in its neighbourhood. As we have already discussed above, the cases $\delta_{1,2}\ll 1$ always imply some tuning of $\delta\lambda$ and $\delta y_t$. Here we do not address the naturalness of the tuning but simply investigate all possible observational signatures. 

We have numerically investigated the inflationary dynamics for each parameter set $(m_h,m_t,\xi,\delta\lambda, \delta y_t)$. We scan over Higgs and top masses within the observational $2$-$\sigma$ bounds $m_h=125.09\pm0.24 $ GeV and $m_t=173.21\pm1.22 $ GeV \cite{PDG} and vary the non-minimal coupling in the range $1\leqslant \xi \leqslant10^4$. We let the matching parameters $\delta\lambda, \delta y_t$ take any values which are consistent with perturbativity $|\lambda(\mu)|<1$ and $|y_t(\mu)|<1$ and satisfy the conditions Eqs. (\ref{successfulreh}), (\ref{Tformula}) and (\ref{negslope}). For each  combination  $(m_h,m_t,\xi,\delta\lambda, \delta y_t)$,  we evaluate the inflationary observables Eq. (\ref{SRobs}) at the field value $\chi_{\rm CMB}$ corresponding to $N_{\rm CMB}  = 60$ efolds before the end of inflation. In this way we are able to systematically determine all the possible observational signatures of the Higgs inflation. Note that by setting $N_{\rm CMB}  = 60$ we neglect the logarithmic dependence on reheating temperature which may differ for different parameter choices. This does not restrict the generality of our results as the scan still covers all possible shapes of the Higgs potential at any inflationary field value. The variation of $N_{\rm CMB}$ would only affect the precise relation between $\delta\lambda$ and $\delta y_t$ and the observational signatures.

The results are depicted in Fig. 1. This shows the possible values of the spectral index $n_s$, the tensor-to-scalar ratio $r_T$ and the running of the spectral index $\alpha_s$ for parameter sets which yield the observed amplitude of perturbations $ {\cal P}_{\zeta}=  (2.139 \pm 0.063) \times 10^{-9}$ \cite{Ade:2015xua}. 
 \begin{figure}[h!]
 \label{fig:nsralpha}
 \begin{center}
\includegraphics[width=0.5\textwidth]{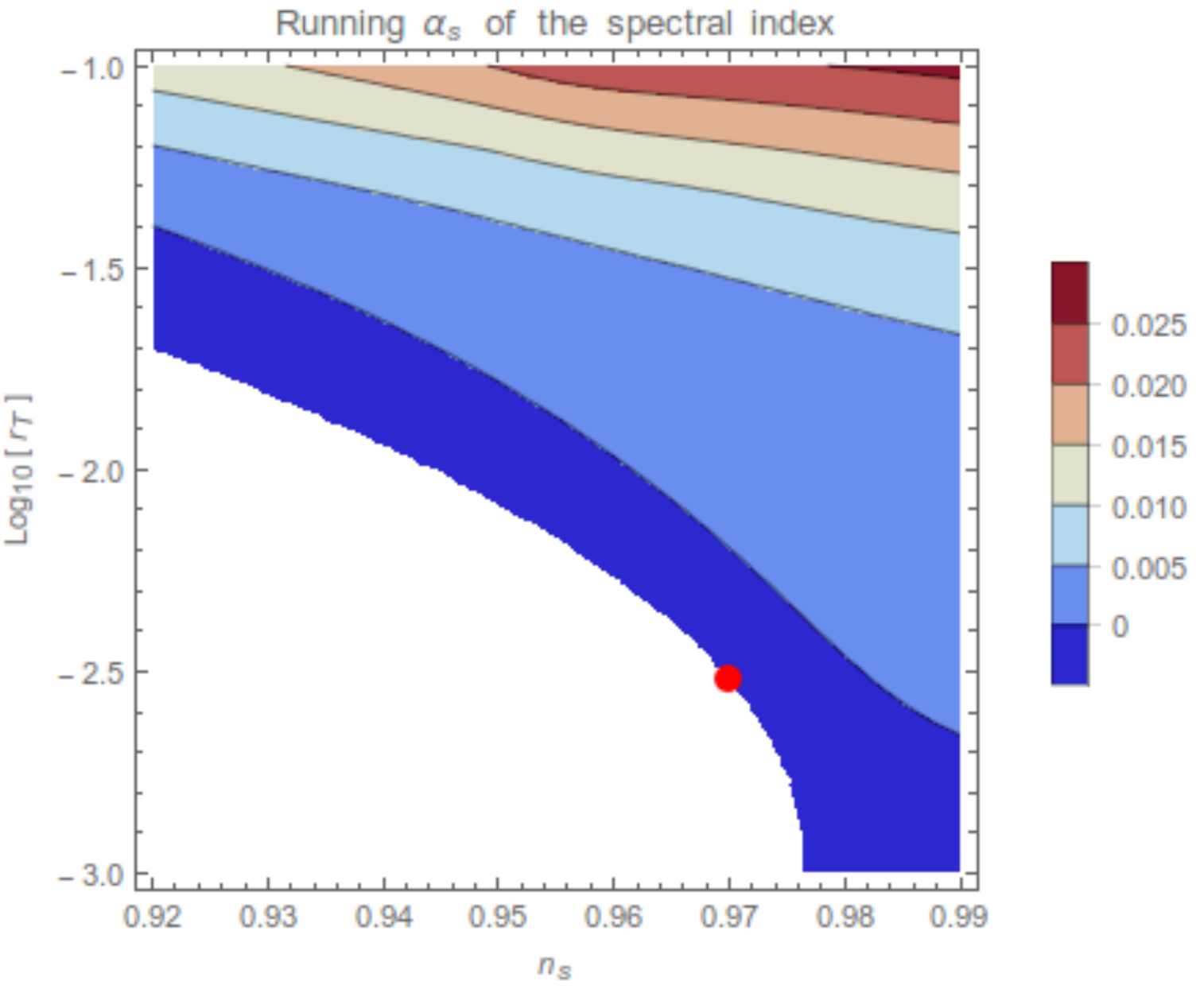}
\caption{The possible combinations of  the spectral index $n_s$, its running $\alpha_s$ and tensor-to-scalar ratio $r_T$ that can be obtained in Higgs inflation requiring that  ${\cal P}_{\zeta}=2.14\times 10^{-9}$. The red point corresponds to the standard case with no features in the inflationary potential. Values in the white regime cannot be obtained in Higgs inflation for any choice of parameters. }
\end{center}
\end{figure}
The standard Higgs inflation, realised for parameter combinations $(m_h,m_t,\xi,\delta\lambda,\delta y_t)$ which yield $\delta_{1,2} = {\cal O}(1)$, corresponds to the red dot in the figure, $n_s= 0.967 $ and $r_T=0.0033$ \cite{Bezrukov:2007ep}. Other points in the plot correspond to matching parameters which yield  $\delta_{1,2}\ll {\cal O}(1)$. The further the point is from the standard prediction, the smaller the values of $\delta_1$ and $\delta_2$ are \footnote{For $\delta_{1} \lesssim 10^{-3}$  the region in the immediate vicinity of the feature may be dominated by quantum fluctuations which lead to essentially stochastic motion for $[V'|< H^3$ \cite{Starobinsky:1994bd}. Removing this range of values for $\delta_1$ does not affect Fig. 1 but the eventual observable effects of the quantum kicks should in principle be investigated separately \cite{Vennin:2015hra}.} Note that as opposed to \cite{Bezrukov:2014ipa} we find parameter combinations  $(m_h,m_t,\xi,\delta\lambda,\delta y_t)$ which yield a large tensor-to-scalar ratio $r_T\gtrsim 0.1$ and still meet the necessary criteria for successful reheating Eqs. (\ref{successfulreh}) and (\ref{Tformula}). They correspond to cases where $\lambda(\mu)$ does not run negative below $\mu_1$ Eq. (\ref{mu1}) and jumps downwards after the matching scale,  $\delta\lambda<0$. 

We stress we have scanned over all possible values of the matching parameters $\delta \lambda$ and $\delta y_t$ consistent with perturbativity and reheating. The results in Fig. 1 therefore show the full range of observational signatures which can be obtained in Higgs inflation for any matching of the low and high energy regimes described by Eq. (\ref{jumps}) (we only show the observationally interesting regime here, the possible values of $n_s$ and $r_T$ do extend outside the depicted range). 

Therefore, we find that even if the matching parameters $\delta \lambda$ and $\delta y_t$ may significantly affect the observables, the Higgs inflation is still falsifiable if only the Lagrangian in the inflationary regime is given by Eq. (\ref{Lnonminimal}).
For example,  a detection of negative running of the spectral index at the level $\alpha_{s}\lesssim -0.01$ would suffice to rule out the model. This holds for any value of the non-minimal coupling $\xi$ and irrespectively of details of the non-renormalisable intermediate scale physics encoded in $\delta \lambda$ and $\delta y_t$. Moreover, even if $r_T$ can be tuned almost independently of $n_s$ by adjusting the parameters $(m_h,m_t,\xi,\delta\lambda,\delta y_t)$, a large tensor-to-scalar ratio within the Higgs inflation is necessarily accompanied by an enhanced positive running of the spectral index as seen in Fig. 1.

\subsection{Observational constraints on the matching}

It is also interesting to investigate in more detail how viable values of  $\delta\lambda$ and $\delta y_t$ within the Higgs inflation scenario depend on SM parameters and the non-minimal coupling $\xi$. 

In the limit of standard Higgs inflation, $\delta_{1,2} = {\cal O}(1)$, the inflationary predictions are given by Eq. (\ref{standardobs}) and do not directly depend on the matching. However, even in this limit the jumps $\delta\lambda$ and $\delta y_t$ cannot be chosen freely. Matching the amplitude of perturbations to the observed value ${\cal P}_{\zeta} = 2.139 \times 10^{-9}$ \cite{Ade:2015xua} and imposing the reheating conditions Eq. (\ref{successfulreh}) heavily constrains the possible choices. The viable values of  $\delta\lambda$ and $\delta y_t$ for the best fit Higgs and top masses are depicted in the left panel of Fig. 2 as a function of the non-minimal coupling $\xi$.  
\begin{figure}[h!]
\begin{center}
\label{fig:toimiva-Inflaatio-SM-best-fit}
\includegraphics[width=0.9\textwidth]{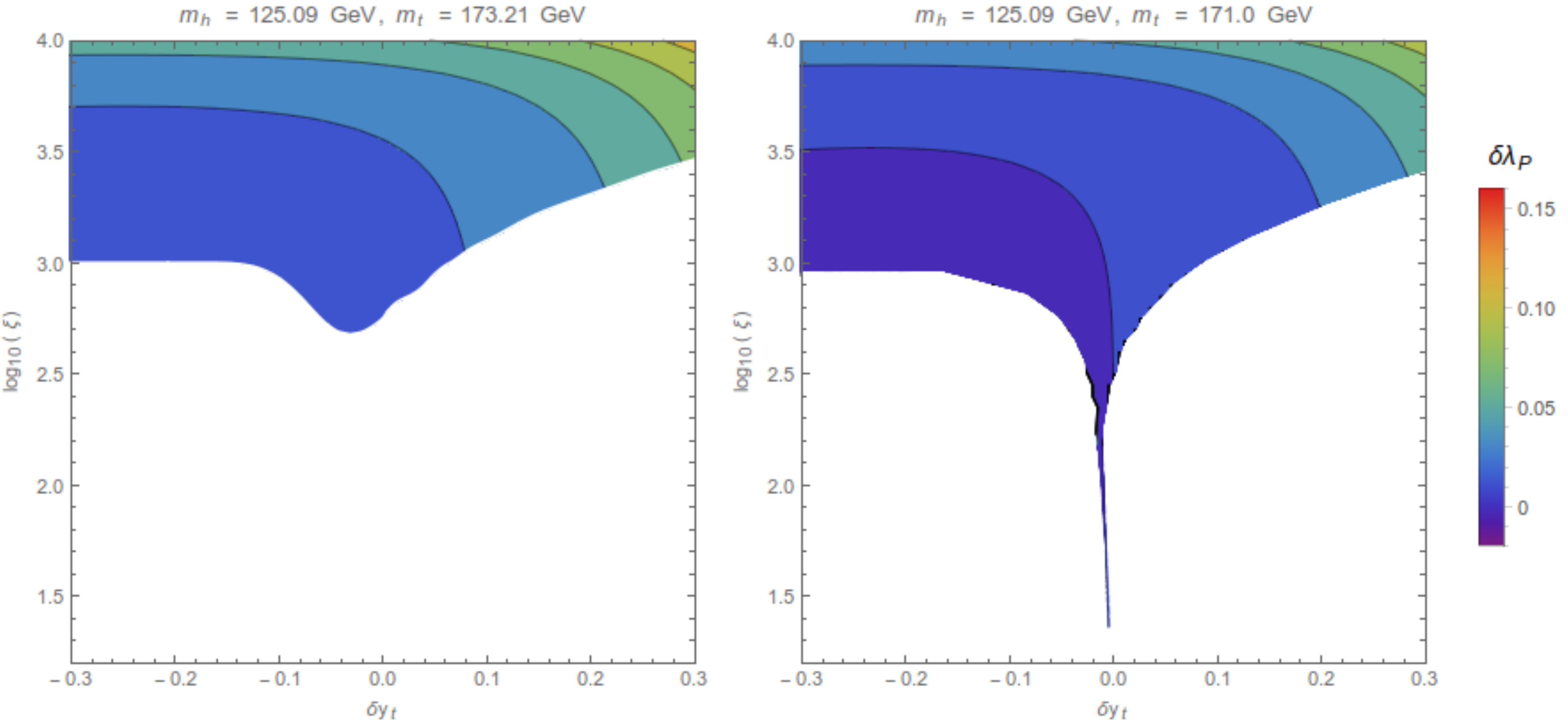}
\caption{The matching parameters $\delta\lambda, \delta y_t$ and non-minimal coupling $\xi$ which yield the observed amplitude of perturbations and allow for reheating into the electroweak vacuum. The left panel shows the results for the best fit values $m_h = 125.09$ GeV, $m_t=173.21$ GeV and the right panel for a small top mass $m_h = 125.09$ GeV, $m_t=170.0$ GeV. In the left panel all points correspond to inflationary potential with no features. In the right panel, points in the regime $\xi\lesssim 300$ yield a feature in the potential.}
\end{center}
\end{figure}
Only positive values of the jump  $\delta\lambda$  are possible in this case as the Higgs coupling $\lambda(\mu)$ runs negative in the SM regime. It is also noteworthy that the required value of the non-minimal coupling sensitively depends on the jumps $\delta\lambda$ and $\delta y_t$. The possible values of the non-minimal coupling are constrained from below by the reheating condition Eq. (\ref{successfulreh}) as the negative minimum gets deeper the smaller the value of $\xi$ is.  However, even for the best fit SM parameters we find that the Higgs inflation can be realised for non-minimal coupling values as low as $\xi={\cal O}(400)$. 

The dependence of the matching parameters on the Higgs and top masses is illustrated in Fig. 3 for different values of $\xi$.  
As the non-minimal coupling decreases, the viable range of $m_h,m_t$ for any choice of matching parameters shrinks towards smaller top masses. This is again due to the reheating condition Eq. (\ref{successfulreh}).  
\begin{figure}[h!]
\begin{center}
\label{fig:pullonpohjat}
\includegraphics[width=0.9\textwidth]{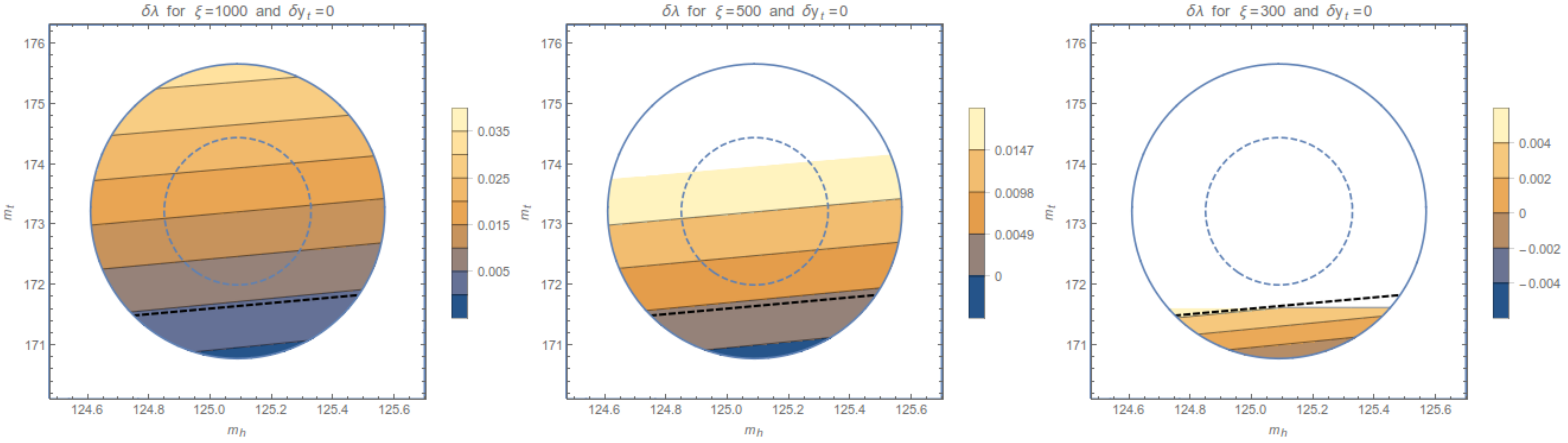}
\caption{The dependence of matching parameters on $m_h$ and $m_t$ in the case with no features in the potential. All points yield the observed amplitude of perturbations and allow for reheating into the electroweak vacuum. Below the dashed line there are also viable choices of $\delta\lambda$ and $\delta y_t$ which generate features in the potential. }
\end{center}
\end{figure}
In Fig.  3 we have only shown the results for matching parameters which yield the standard Higgs inflation with no features $\delta_{1,2} = {\cal O}(1)$. For $m_t\gtrsim 171.8$ GeV, the choices which would generate features in the potential always yield a too small amplitude for perturbations.  
This is because the amplitude scales as ${\cal P}_{\zeta}\propto \xi^{-2}$ and for $m_t\gtrsim 171.8$ GeV the reheating condition Eq.  (\ref{successfulreh}) constrains $\xi$ to values which are too large to yield the observed amplitude for $\delta_{1,2}\ll 1$. The predictions of Higgs inflation are thus uniquely given by Eq. (\ref{standardobs}) for top mass values in the range  $m_t\gtrsim 171.8$ GeV.

The situation is different in the regime $m_t\lesssim 171.8$ GeV marked by the dashed line in Fig. 3. Here the observed amplitude of perturbations can be obtained also for $\delta\lambda$ and $\delta y_t$ which generate a feature in the potential, $\delta_{1,2}\ll 1$, and still pass the reheating condition Eq. (\ref{successfulreh}). This is also seen in the right panel of Fig. 2 which illustrates the possible values of $\delta\lambda, \delta y_t$ and $\xi$ for a successful Higgs inflation with $m_t=171.0$ GeV.  In the regime $\xi \gtrsim 300 $, the viable values of $\delta\lambda$ and $\delta y_t$ yield $\delta_{1,2}= {\cal O}(1)$ and lead to the standard case with no features.  The region $\xi \lesssim 300$ on the other hand corresponds to matching parameters which generate a feature in the potential, $\delta_{1,2}\ll 1$. This regime is absent in the left panel which shows the corresponding results for the best fit values of SM parameters. In this case the observed amplitude can only be obtained in the regime with no features. 

In the presence of a feature the observational signatures are not unique but depend on the matching and vary over the values shown in Fig. 1. Changing $\delta\lambda$ and $\delta y_t$ affects both the shape of the feature (the values of $\delta_{1,2}$ in Eqs. (\ref{delta1}) and (\ref{delta2})) and its location compared to the field value $\chi_{\rm CMB}$ at the horizon crossing of observable modes. The dependence on $\delta\lambda$ and $\delta y_t$ is in general different for different choices of $m_h,m_t$ and $\xi$.\begin{figure}[h!]
\begin{center}
\label{fig:dlnsplot}
\includegraphics[width=0.9 \textwidth]{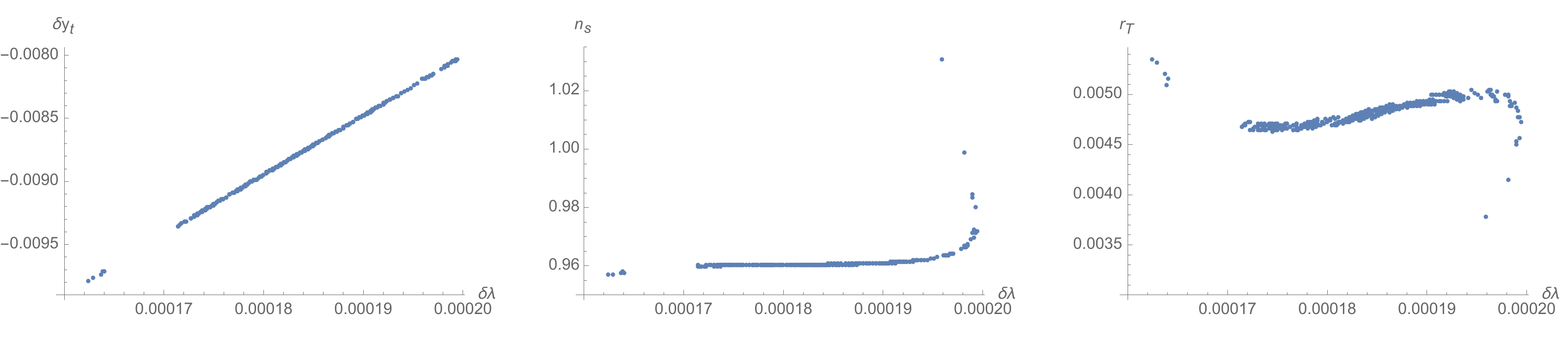}
\caption{The dependence of the spectral index $n_s$ and tensor-to-scalar ratio $r_T$ on the matching parameters for a sample point $m_h = 125.09$ GeV, $m_t=171.05$ GeV and $\xi =59$. All points yield the observed amplitude of perturbations and the mutual dependence of $\delta \lambda$ and $\delta y_t$ is shown in left panel.}
\end{center}
\end{figure}
An example is seen in Fig. 4 which shows how the spectral index and tensor to scalar ratio depend on the matching parameters for a sample point $m_h = 125.09$ GeV, $m_t = 171.05$ GeV and $\xi = 59$.

The smallest possible values of the non-minimal coupling $\xi$ for which the observed amplitude can be obtained are depicted in Fig. 5. All points shown in the figure pass the minimal conditions for successful reheating  Eq. (\ref{successfulreh}). As already noticed in \cite{Bezrukov:2014bra,Bezrukov:2014ipa} the Higgs inflation can be realised even for $\xi \lesssim 10$  in the regime of low top mass values. The viable choices of matching parameters  $\delta\lambda$ and $\delta y_t$  corresponding to small $\xi$ all yield features in the inflationary potential.
\begin{figure}[h!]
\begin{center}
\includegraphics[width=0.5 \textwidth]{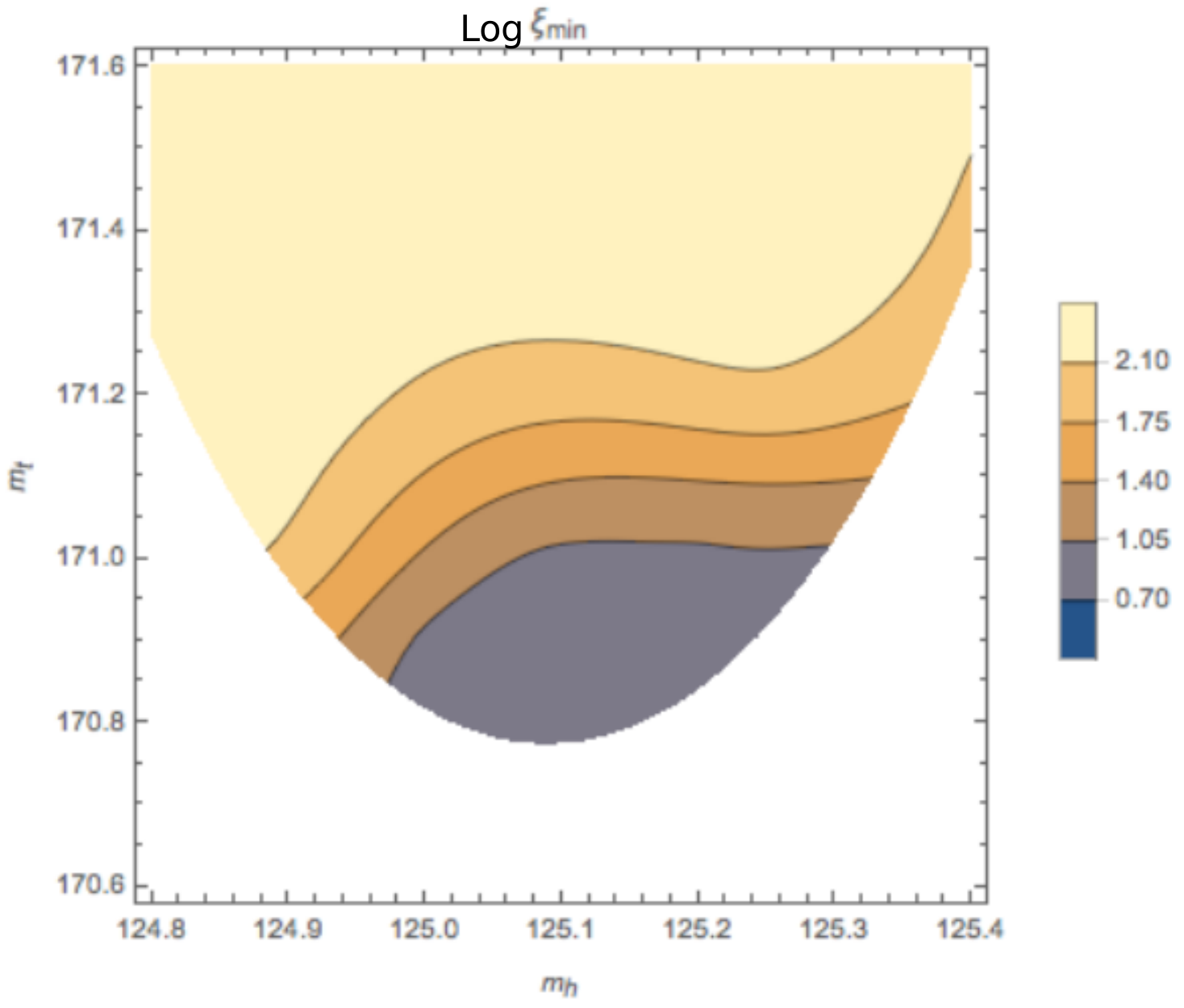}
\caption{The smallest possible values of the non-minimal coupling $\xi$ for which the observed amplitude of perturbations can be obtained and reheating into the electroweak vacuum is possible.}
\end{center}
\end{figure}

\section{Conclusions}
\label{sec:conclusions}

In Higgs inflation the connection between the SM limit at low energies and the approximatively shift symmetric inflationary plateau at high energies is obscured by manifestly non-renormalisable physics at intermediate scales. A possible resolution is that the non-minimally coupled inflationary limit is obtained as a UV completion of SM coupled to gravity. Without the exact solution at hand, however, the matching of the effective low and high energy limits in the scenario is to large extent ambiguous.

Here we have thoroughly analysed the observational signatures of Higgs inflation and their dependence on the mathcing. Following \cite{Bezrukov:2014ipa} we model the impacts of unknown non-renormalisable physics by two free parameters $\delta \lambda$ and $\delta y_t$ which connect the low and high energy running of Higgs and top Yukawa couplings. By varying the Higgs and top masses within their observational two sigma bounds, letting the non-minimal coupling to take values in the range $\xi=1...10^4$ and scanning over $\delta \lambda$ and $\delta y_t$ we have established a systematical picture of the model predictions. 

We find that generating the observed amplitude of perturbations in general requires non-zero positive values for $\delta \lambda$. For example, for the best fit values $m_t = 173.21$ GeV and $m_h = 125.09$ GeV the Higgs coupling runs negative at $\mu \simeq 10^{11}$ GeV and a jump $\delta \lambda\sim 0.01$ to positive values is necessary for inflation to take place. Both $\delta \lambda$ and $\delta y_t$ and the non-minimal coupling $\xi$ are further constrained by the required decay of the inflationary Higgs condensate into the electroweak vacuum despite the negative energy minimum in the zero temperature potential.  Perturbations and the reheating constraint together make the required $\xi$ value dependent on $\delta \lambda$ and $\delta y_t$ and even relatively small values of $\xi$ can lead to successful inflation. For the best fit SM parameters, the lowest possible value is $\xi \simeq 400$. 

For top mass values in the range $m_t\gtrsim 171.8$ GeV we find that successful Higgs inflation is only possible for matching parameters which yield the standard $R^2$ inflation predictions for the spectral index $n_s\simeq 0.97$ and tensor-to-scalar ratio $r_T\simeq 0.003$. In the regime $m_t\lesssim 171.8$ GeV the observed amplitude of perturbations together with successful reheating can also be obtained for $\delta \lambda$ and $\delta y_t$ which are tuned to generate a feature in the inflationary potential. In this case, $n_s$ and $r_T$ can vary over broad range depending on the values of the matching parameters. Remarkably, however, we find that the scenario is falsifiable even in this limit. For any choices of $\delta \lambda, \delta y_t$ and $\xi$ the predictions for $n_s$,  its running $\alpha_s$ and $r_T$ are constrained to the surface shown in Fig. 1. For example, a detection of negative running of the spectral index at level $\alpha_s\lesssim -0.01$ would rule out the scenario irrespectively of details of the matching.

\acknowledgments{V-M.E. is supported by the Magnus Ehrnrooth foundation.}



\begin{thebibliography}{7}

 
\bibitem{Bezrukov:2007ep}
  F.~L.~Bezrukov and M.~Shaposhnikov,
  Phys.\ Lett.\ B {\bf 659} (2008) 703
  doi:10.1016/j.physletb.2007.11.072
  [arXiv:0710.3755 [hep-th]].
 
 
 \bibitem{Burgess:2014lza}
  C.~P.~Burgess, S.~P.~Patil and M.~Trott,
  JHEP {\bf 1406} (2014) 010
  doi:10.1007/JHEP06(2014)010
  [arXiv:1402.1476 [hep-ph]];
  C.~P.~Burgess, H.~M.~Lee and M.~Trott,
  JHEP {\bf 0909} (2009) 103
  doi:10.1088/1126-6708/2009/09/103
  [arXiv:0902.4465 [hep-ph]];
  C.~P.~Burgess, H.~M.~Lee and M.~Trott,
  JHEP {\bf 1007} (2010) 007
  doi:10.1007/JHEP07(2010)007
  [arXiv:1002.2730 [hep-ph]];
  J.~L.~F.~Barbon and J.~R.~Espinosa,
  Phys.\ Rev.\ D {\bf 79} (2009) 081302
  doi:10.1103/PhysRevD.79.081302
  [arXiv:0903.0355 [hep-ph]].

 
 \bibitem{Bezrukov:2010jz}
  F.~Bezrukov, A.~Magnin, M.~Shaposhnikov and S.~Sibiryakov,
  JHEP {\bf 1101} (2011) 016
  doi:10.1007/JHEP01(2011)016
  [arXiv:1008.5157 [hep-ph]].

\bibitem{Bezrukov:2014bra}
  F.~Bezrukov and M.~Shaposhnikov,
  Phys.\ Lett.\ B {\bf 734} (2014) 249
  doi:10.1016/j.physletb.2014.05.074
  [arXiv:1403.6078 [hep-ph]].

\bibitem{Bezrukov:2014ipa}
  F.~Bezrukov, J.~Rubio and M.~Shaposhnikov,
  Phys.\ Rev.\ D {\bf 92} (2015) 8,  083512
  doi:10.1103/PhysRevD.92.083512
  [arXiv:1412.3811 [hep-ph]].
  
  
  
\bibitem{Fumagalli:2016lls}
  J.~Fumagalli and M.~Postma,
  arXiv:1602.07234 [hep-ph].
  
\bibitem{Bezrukov:2009db}
  F.~Bezrukov and M.~Shaposhnikov,
  JHEP {\bf 0907} (2009) 089
  doi:10.1088/1126-6708/2009/07/089
  [arXiv:0904.1537 [hep-ph]];
  F.~L.~Bezrukov, A.~Magnin and M.~Shaposhnikov,
  Phys.\ Lett.\ B {\bf 675} (2009) 88
  doi:10.1016/j.physletb.2009.03.035
  [arXiv:0812.4950 [hep-ph]].
  
  \bibitem{George:2013iia}
  D.~P.~George, S.~Mooij and M.~Postma,
  JCAP {\bf 1402} (2014) 024
  doi:10.1088/1475-7516/2014/02/024
  [arXiv:1310.2157 [hep-th]];
  D.~P.~George, S.~Mooij and M.~Postma,
  arXiv:1508.04660 [hep-th].


\bibitem{Barbon:2015fla}
  J.~L.~F.~Barbon, J.~A.~Casas, J.~Elias-Miro and J.~R.~Espinosa,
  JHEP {\bf 1509} (2015) 027
  doi:10.1007/JHEP09(2015)027
  [arXiv:1501.02231 [hep-ph]].

  
  \bibitem{Degrassi}

  G.~Degrassi, S.~Di Vita, J.~Elias-Miro, J.~R.~Espinosa, G.~F.~Giudice, G.~Isidori and A.~Strumia,
  JHEP {\bf 1208} (2012) 098
  [arXiv:1205.6497 [hep-ph]];
  D.~Buttazzo, G.~Degrassi, P.~P.~Giardino, G.~F.~Giudice, F.~Sala, A.~Salvio and A.~Strumia;
  JHEP {\bf 1312} (2013) 089
  [arXiv:1307.3536].
  

  
  
  \bibitem{Hamada:2014iga}
  Y.~Hamada, H.~Kawai, K.~y.~Oda and S.~C.~Park,
  Phys.\ Rev.\ Lett.\  {\bf 112} (2014) no.24,  241301
  doi:10.1103/PhysRevLett.112.241301
  [arXiv:1403.5043 [hep-ph]].


\bibitem{Fairbairn:2014nxa}
  M.~Fairbairn, P.~Grothaus and R.~Hogan,
  JCAP {\bf 1406} (2014) 039
  doi:10.1088/1475-7516/2014/06/039
  [arXiv:1403.7483 [hep-ph]];
  I.~Masina,
  Phys.\ Rev.\ D {\bf 89} (2014) no.12,  123505
  doi:10.1103/PhysRevD.89.123505
  [arXiv:1403.5244 [astro-ph.CO]];
  J.~L.~Cook, L.~M.~Krauss, A.~J.~Long and S.~Sabharwal,
  Phys.\ Rev.\ D {\bf 89} (2014) no.10,  103525
  doi:10.1103/PhysRevD.89.103525
  [arXiv:1403.4971 [astro-ph.CO]];
  C.~Germani, Y.~Watanabe and N.~Wintergerst,
  JCAP {\bf 1412} (2014) no.12,  009
  doi:10.1088/1475-7516/2014/12/009
  [arXiv:1403.5766 [hep-ph]].


  
 \bibitem{Ade:2015xua}
  P.~A.~R.~Ade {\it et al.} [Planck Collaboration],
  arXiv:1502.01589 [astro-ph.CO].

\bibitem{Barvinsky:2008ia}
  A.~O.~Barvinsky, A.~Y.~Kamenshchik and A.~A.~Starobinsky,
  JCAP {\bf 0811} (2008) 021
  doi:10.1088/1475-7516/2008/11/021
  [arXiv:0809.2104 [hep-ph]];
  A.~De Simone, M.~P.~Hertzberg and F.~Wilczek,
  Phys.\ Lett.\ B {\bf 678} (2009) 1
  doi:10.1016/j.physletb.2009.05.054
  [arXiv:0812.4946 [hep-ph]];
  A.~O.~Barvinsky, A.~Y.~Kamenshchik, C.~Kiefer, A.~A.~Starobinsky and C.~Steinwachs,
  JCAP {\bf 0912} (2009) 003
  doi:10.1088/1475-7516/2009/12/003
  [arXiv:0904.1698 [hep-ph]];
  A.~O.~Barvinsky, A.~Y.~Kamenshchik, C.~Kiefer, A.~A.~Starobinsky and C.~F.~Steinwachs,
  Eur.\ Phys.\ J.\ C {\bf 72} (2012) 2219
  doi:10.1140/epjc/s10052-012-2219-3
  [arXiv:0910.1041 [hep-ph]].



\bibitem{Espinosa:2007qp}
  J.~R.~Espinosa, G.~F.~Giudice and A.~Riotto,
  JCAP {\bf 0805} (2008) 002
  doi:10.1088/1475-7516/2008/05/002
  [arXiv:0710.2484 [hep-ph]];


  A.~Kobakhidze and A.~Spencer-Smith,
  Phys.\ Lett.\ B {\bf 722} (2013) 130
  [arXiv:1301.2846 [hep-ph]];

  M.~Fairbairn and R.~Hogan,
  Phys.\ Rev.\ Lett.\  {\bf 112} (2014) 201801
  [arXiv:1403.6786 [hep-ph]];

  K.~Enqvist, T.~Meriniemi and S.~Nurmi,
  JCAP {\bf 1407} (2014) 025
  [arXiv:1404.3699 [hep-ph]];

  A.~Hook, J.~Kearney, B.~Shakya and K.~M.~Zurek,
  JHEP {\bf 1501} (2015) 061
  [arXiv:1404.5953 [hep-ph]];

  J.~R.~Espinosa, G.~F.~Giudice, E.~Morgante, A.~Riotto, L.~Senatore, A.~Strumia and N.~Tetradis,
  JHEP {\bf 1509} (2015) 174
  doi:10.1007/JHEP09(2015)174
  [arXiv:1505.04825 [hep-ph]].

\bibitem{Herranen:2014cua}
  M.~Herranen, T.~Markkanen, S.~Nurmi and A.~Rajantie,
  Phys.\ Rev.\ Lett.\  {\bf 113} (2014) 21,  211102
  [arXiv:1407.3141 [hep-ph]];
  
  M.~Herranen, T.~Markkanen, S.~Nurmi and A.~Rajantie,
  Phys.\ Rev.\ Lett.\  {\bf 115} (2015) 241301
  doi:10.1103/PhysRevLett.115.241301
  [arXiv:1506.04065 [hep-ph]];

  I.~G.~Moss,
  arXiv:1509.03554 [hep-th].


\bibitem{Lebedev}

  O.~Lebedev,
  Eur.\ Phys.\ J.\ C {\bf 72} (2012) 2058
  [arXiv:1203.0156 [hep-ph]];


  O.~Lebedev and A.~Westphal,
  Phys.\ Lett.\ B {\bf 719} (2013) 415
  [arXiv:1210.6987 [hep-ph]].



\bibitem{Dutta:2007st}
  S.~Dutta, K.~Hagiwara, Q.~S.~Yan and K.~Yoshida,
  Nucl.\ Phys.\ B {\bf 790} (2008) 111
  doi:10.1016/j.nuclphysb.2007.08.017
  [arXiv:0705.2277 [hep-ph]].
 
 \bibitem{bezrukoodi}
 http://www.inr.ac.ru/~fedor/SM/


\bibitem{Bezrukov:2008ut}
  F.~Bezrukov, D.~Gorbunov and M.~Shaposhnikov,
  JCAP {\bf 0906} (2009) 029
  doi:10.1088/1475-7516/2009/06/029
  [arXiv:0812.3622 [hep-ph]];
  J.~Garcia-Bellido, D.~G.~Figueroa and J.~Rubio,
  Phys.\ Rev.\ D {\bf 79} (2009) 063531
  doi:10.1103/PhysRevD.79.063531
  [arXiv:0812.4624 [hep-ph]].


\bibitem{Enqvist:2015sua}
  K.~Enqvist, S.~Nurmi, S.~Rusak and D.~Weir,
  JCAP {\bf 1602} (2016) no.02,  057
  doi:10.1088/1475-7516/2016/02/057
  [arXiv:1506.06895 [astro-ph.CO]];
  D.~G.~Figueroa, J.~Garcia-Bellido and F.~Torrenti,
  Phys.\ Rev.\ D {\bf 92} (2015) no.8,  083511
  doi:10.1103/PhysRevD.92.083511
  [arXiv:1504.04600 [astro-ph.CO]];
  K.~Enqvist, S.~Nurmi and S.~Rusak,
  JCAP {\bf 1410} (2014) no.10,  064
  doi:10.1088/1475-7516/2014/10/064
  [arXiv:1404.3631 [astro-ph.CO]].

\bibitem{Kajantie:1995dw}
  K.~Kajantie, M.~Laine, K.~Rummukainen and M.~E.~Shaposhnikov,
  Nucl.\ Phys.\ B {\bf 458} (1996) 90
  doi:10.1016/0550-3213(95)00549-8
  [hep-ph/9508379].

\bibitem{Starobinsky:1980te}
  A.~A.~Starobinsky,
  Phys.\ Lett.\ B {\bf 91} (1980) 99.
  doi:10.1016/0370-2693(80)90670-X


\bibitem{PDG}
K.A. Olive et al. (Particle Data Group), Chin. Phys. C, 38, 090001 (2014) and 2015 update.

\bibitem{Starobinsky:1994bd}
  A.~A.~Starobinsky and J.~Yokoyama,
  Phys.\ Rev.\ D {\bf 50} (1994) 6357
  doi:10.1103/PhysRevD.50.6357
  [astro-ph/9407016].

\bibitem{Vennin:2015hra}
  V.~Vennin and A.~A.~Starobinsky,
  Eur.\ Phys.\ J.\ C {\bf 75} (2015) 413
  doi:10.1140/epjc/s10052-015-3643-y
  [arXiv:1506.04732 [hep-th]].


\end{thebibliography}
\end{document}